# Bypassing Content-based internet packages with an SSL/TLS Tunnel, SNI Spoofing, and DNS spoofing

Shanaka Anuradha Samarakoon
Freelancer
shanaka95@gmail.com

*Abstract*— Internet Service Providers (ISPs) are increasingly offering content-based packages to their clients. These packages offer access to a range of online content, such as Facebook, YouTube, Messenger, Zoom, and many other popular services, for a fixed price. This allows users to access all the content they want without worrying about data caps or overage charges. These packages are way cheaper than regular internet packages. Even some ISPs offer unlimited content-based packages for a low price. When using these packages, network traffic is continuously filtered by the ISP, and the user will be charged separately for using other services which are not included in the content-based package.

Some internet users are using HTTP injector Software to bypass ISP's Network traffic filters and access other resources available on the internet using content-based package data quotes. This research aims to find an alternative method to bypass ISP's Network traffic filters without using an HTTP injector.

## I. INTRODUCTION

HTTP injectors are tools used to fool the internet service provider and make other web resources appear as if they're completely free. This allows the user to access content that would otherwise be blocked by the package filter. An example would be to visit a website such as facebook.com by using a content-based internet package that only allows certain websites to be accessed, such as Zoom, Google Meet, and Microsoft Teams.

There are significant disadvantages of using HTTP injectors to bypass ISP's Network traffic filters.

- HTTP Injector software might include malware
- SSL/TSL enabled SSH account is required.
- Free SSH accounts do not guarantee Uptime and Connection Stability.
- Free SSH accounts do not guarantee Security and Privacy.
- Free SSH accounts expire in a few hours.
- Free SSH accounts offer limited connection speed.

This research paper aims to find another way to spoof the ISP and surf the internet without using HTTP Injectors. We need to use a traffic filter bypassing technique to spoof ISPs' traffic filters. There are many types of traffic filters used in the IT and telecommunications industry. Therefore, before trying to bypass them, we need to identify which type of filters are being used. Therefore, the first objective of this research study is to identify what type of Traffic Filters are used by ISPs to limit users from surfing websites other than those included ones in the content-based package.

The second objective of this research study is to develop a methodology that can be used to utilize any resource available on the internet by using a content-based package data quota. This proposed methodology should be able to configure on a normal user computer effortlessly. As the final objective of this study, it is required to test the proposed methodology in a typical user computing environment to ensure its feasibility.

To set up the testing environment for this study, I have used a Linux Desktop PC as the user computer and another Linux VPS instance as a remote server. This setup allowed me to have full control over the testing environment, which was essential for getting accurate results. More importantly, I have used Sri Lankan ISP SLT Mobitel with Learn from the home package, which includes a 100GB Data Quote for the Zoom application.

According to internetsociety.org [1], four possible filtering techniques are available to filter network traffic.

- IP and Protocol-based filtering
- Deep Packet Inspection-based filtering
- URL-based filtering
- Platform-based filtering
- DNS-based filtering

As discussed by W. M. Shbair, T. Cholez et al. [2], websites are increasingly moving to use HTTPS to provide a secure connection for users. HTTPS encrypts the whole data packet, making it impossible for ISPs to inspect data packets or identify the website's hostname. This prevents ISPs from being able to throttle or block specific content on the web. Due to this reason, ISPs use SNI-based traffic filters to filter HTTPS web traffic.

Users can easily bypass Deep Packet Inspection-based, URL-based, and Platform-based filters using an HTTPS enabled Virtual private network (VPN). VPN encrypts all the data transferred through the network, making it almost impossible to filter traffic by Data packets, URLs, or Platforms.

There are three possible techniques available that Internet Service Providers (ISPs) can use to block or filter traffic transferred through VPN servers or HTTPS. One way is to use DNS-based filters. This type of filter can block traffic to specific domains or domain names. ISPs can block or restrict VPN servers or other websites by configuring their DNS servers to return the wrong IP addresses for domain names. But internet users can easily move to other freely available DNS servers from third-party providers and bypass the DNS blocking.

Another way is to use IP and Protocol-based filters. These filters can be used to block traffic to specific IP addresses or to specific protocols. Also, SNI (Server Name Identification) based web filtering can be used to block access to specific websites or to allow access to only specific websites. This paper presents a detailed study on techniques that we can use to bypass IP and Protocol-based traffic filters, DNS-based traffic filters, and SNI-based traffic filters to access any resource available on the internet using content-based internet packages.

## II. METHODOLOGY

Traffic filter bypassing can be achieved, if we are able to route all web traffic to one of our own VPS instance, convincing the ISP that we are connecting to an allowed web server such as zoom.us webserver. After achieving this, all traffic can be re-routed to intended destinations through a proxy server or VPN installed inside our VPS instance. This should allow the users to use content-based package data quotes. However, convincing ISPs is not simply because they are using multiple traffic filters at once.

### A. Traffic Filtering

T Usually, Zoom, Facebook, WhatsApp, Messenger, and most other services use a wide range of IP addresses on their servers. Also, they use CDNs to provide their service worldwide. Due to this reason, ISPs cannot filter user traffic by destination IP address. Because they do not know the static set of IP addresses related to the content-based package. Therefore, ISPs use domain names to filter the network traffic.

As discussed by D. Storm [2], Even with HTTPS, ISPs can still see the domains that their subscribers visit (Figure 1). ISPs analyze DNS queries to identify websites that users are visiting. To prevent this, we can get rid of using ISP's DNS server and spoof DNS queries as mentioned in the below sections.

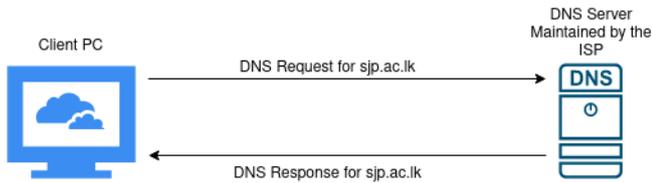

Figure 1. Resolving DNS queries by the ISP

W. M. Shbair, T. Cholez et al. [7] have discussed another approach for filtering HTTPS-based web traffic. The Server Name Indication (SNI) technique is used by many firewalls and content-filtering solutions to filter HTTPS traffic, as they discussed in their research study. Figure 2 demonstrates how SNI details are stored inside HTTPS header data.

Figure 2. How SNI details are stored inside HTTPS Headers

For this study, we have focused on DNS Spoofing and SNI Spoofing to bypass traffic filters.

### B. DNS Spoofing

DNS spoofing is an attack that allows an attacker to redirect traffic intended for one website domain to another website, usually for malicious purposes. DNS spoofing is often used in conjunction with other types of attacks, such as phishing or malware, to trick users into providing sensitive information or downloading malicious software. DNS spoofing can also be used to redirect traffic away from a website that is under attack, such as a denial-of-service attack. But here in this study, our focus is to trick the ISP, not the user.

Figure 3. The host name is visible to the ISP with an HTTP request.

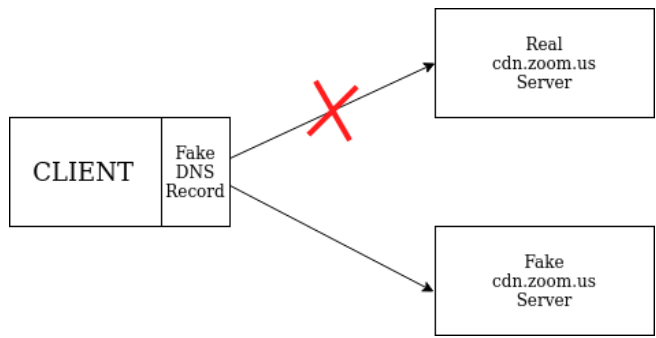

Figure 4. DNS Spoofing Client Server Architecture Diagram

As shown in Figure 3, DNS Spoofing should work with HTTP traffic because ISPs can inspect HTTP data packets and determine the hostname of the destination server. Since we have done DNS spoofing in the local computer as shown in figure 4, data packets will contain the spoofed domain name as the hostname. But when it comes to HTTPS traffic, the encryption prevents the ISP from looking into the data packets and determining the hostname. Therefore, the ISP can't determine the destination server, and DNS Filtering will not work for the ISP. Therefore, DNS Spoofing will not work for HTTPS connections. However, some new techniques are being developed that allow ISPs to inspect HTTPS data packets and determine the hostname of the destination server. One such technique is called SNI (Server Name Indication).

### C. SNI (Server Name Indication)

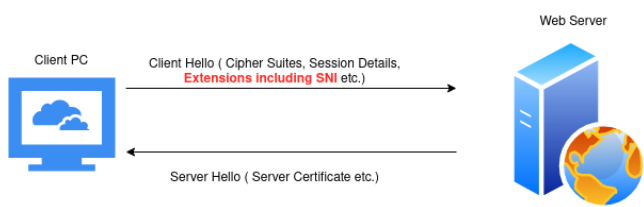

Figure 5. The host name is visible to the ISP with an HTTP request.

When initiating a TCP connection with the server, SNI details are sent in the Client Hello message, as shown in Figure 5. The server responds with the appropriate certificate based on the SNI details included in the Client Hello message. If the SNI details are not included in the Client Hello message, the server responds with the default certificate. With SNI, the hostname of the destination server is included in the data packets in plain text (Figure 2). This allows the ISP to inspect the data packets and determine the hostname of the destination server. SNI-based web filtering is a new technique that is being developed to allow ISPs to inspect HTTPS data packets and determine the hostname of the destination server.

### D. Creating a Web Server with a Fake Domain to spoof SNI

Creating a web server with a fake domain can be a useful way to trick the ISP. For this study, I have used a content-based package that has access to zoom services. Therefore, I have altered my local computer DNS records to point the domain name cdn.zoom.us to one of my remote Linux VPS instances, which has the IP address 178.120.118.221 and an apache web server installed in it.

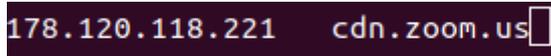

Figure 6. Fake DNS Record on Local Computer

Due to this DNS spoofing technique, All the network traffic from my local computer to cdn.zoom.us will contain the hostname and the SNI of cdn.zoom.us, but they will be routed to the fake server.

### E. Accessing the Fake Webserver

When accessing http://cdn.zoom.us from the web browser, ignoring the SSL not available warning, it brought me to the apache server, which was configured on the VPS.

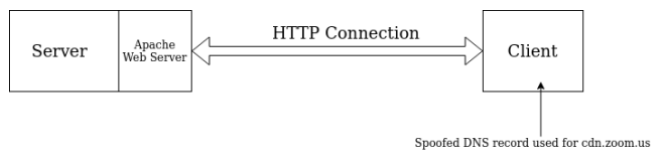

Figure 7. DNS Spoofing without SSL, Client Server Architecture Diagram

### F. Configuring SSL on the Fake Web Server

The next step was to try the above-mentioned process with an SSL-enabled connection. To achieve that, a self-signed certificate was generated for cdn.zoom.us on the same VPS instance, and the Apache webserver was configured to use SSL (Figure 8). Then the SSL CA was added to the local computer as a trusted CA. This step is required since the SSL certificate is not generated by a trusted CA.

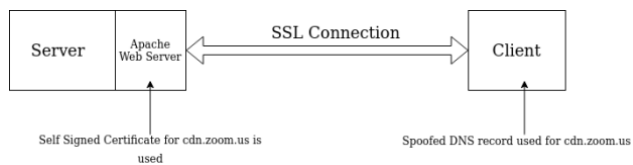

Figure 8. DNS Spoofing with SSL, Client Server Architecture Diagram

### G. Creating an SSL/TLS Tunnel using Stunnel

I continued the research study on initiating an SSL-enabled connection between my Local PC and my remote server while making it impossible for the ISP to track the real destination of data packets. An SSL/TLS tunnel is a secure communication channel between two computers or devices. SSL/TLS tunnels use the SSL/TLS protocol to encrypt and decrypt data. Tunnels can be used to protect data in transit or to create a secure connection between two computers or devices.

SSL/TLS tunnels are often used to secure communications between a web browser and a web server. However, they can also be used to secure other types of traffic, such as email, file transfer, and VPN connections. SSL/TLS tunnels provide a high level of security and are often used in situations where data privacy is of paramount importance. However, they can also be slower than other communication channels due to the overhead associated with encryption and decryption.

Stunnel is free and open-source software that provides a secure tunneling service for internet traffic. It can be used to secure communications between two computers or tunnel traffic through a network. Stunnel uses the SSL/TLS protocol to encrypt traffic and can be used with a variety of protocols, including HTTP, SMTP, POP3, and IMAP. Stunnel is available for Windows, Linux, and macOS. [3]

I configured the Stunnel server application on the VPS instance using a self-signed certificate for the hostname cdn.zoom.us. Then the Certificate CA was added to the local computer as a Trusted CA. Finally, the stunnel client was installed on the local machine using another self-signed certificate. The process mentioned above can be used to initiate a secure SSL/TLS tunnel between the remote server and the local computer.

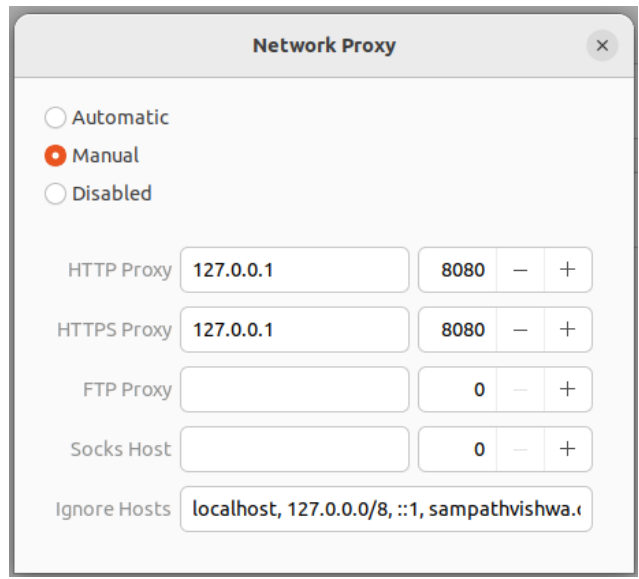

Figure 11. Configuration of System Proxy

![Figure 9]

Figure 9. Configuration of Stunnel Server Application

As shown in Figure 9, Stunnel Server Application is configured to use self-signed certificates which include the hostname as cdn.zoom.us. It is accepting connections on port 443, which is the default port for SSL connections. Finally, all traffic coming through the tunnel is re-routed to the application listening at local port 3128. As mentioned in the below sections, Port 3128 is being used by Squad proxy which can re-route received data to original destinations.

![Figure 10]

Figure 10. Configuration of Stunnel Client Application

As shown in Figure 10, the Stunnel client application is configured on the local PC. The application accepts data through the local port 8080 and then sends the data through the tunnel connection established with cdn.zoom.us:443.

As the final step, the Proxy settings for the Local PC were changed to use port 8080 as shown in Figure 11. This re-routes all network traffic to the local port 8080, which is used by the Stunnel Application.

### H. Configuring Squid proxy

A proxy server is a server that acts as an intermediary between a client and another server. A squid proxy is a specific type of proxy server designed to improve the performance of web applications that use the HTTP protocol[4]. Squid proxies can cache frequently accessed web pages and files, which can significantly improve the performance of web applications. In addition, squid proxies can compress web traffic, which can also improve performance.

To handle the web traffic coming through Stunnel, The squid proxy server was installed on the remote server and was configured to use TCP port 3128. I then configured the Stunnel server application to connect with the TCP port 3128 as mentioned in the previous section. Therefore, the traffic coming through the SSL/TSL tunnel will be routed to the Squid proxy server. In the previous section, I mentioned that all web traffic is now being re-routed through stunnel tunnel to the Squid proxy application installed in the server. Squid proxy then re-routes the web traffic to the original destination. Figure 12 illustrates the overall configuration of the system

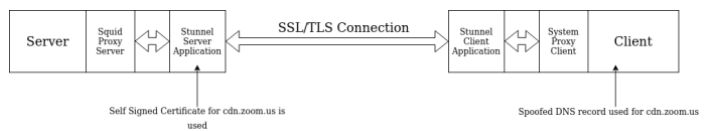

Figure 12. Stunnel and Squad Proxy, Client Server Architecture Diagram

### I. Configuring OpenVPN

I did some additional testing with a VPN application to see if the configuration I mentioned earlier would work. OpenVPN is a free and open-source software application that implements virtual private network (VPN) techniques to create secure point-to-point or site-to-site connections in routed or bridged configurations and remote access facilities. It uses a custom security protocol that utilizes SSL/TLS for key exchange. It is capable of traversing network address translators (NATs) and firewalls. OpenVPN has been ported to multiple platforms, including Linux, Windows, Mac OS X, FreeBSD, NetBSD, OpenBSD, Solaris, Android, iOS, and Windows Phone.

I installed the OpenVPN server on my remote VPN server and was configured to use TCP port 1194. Then I changed Stunnel configurations to connect to the TCP port 1194. The Stunnel client application is still listening on the TCP port 8080 of the local computer. Then I downloaded the OpenVPN client config file to the local computer and changed the connection hostname to 127.0.0.1 and port to 8080. Using the OpenVPN client application, the local computer was connected to the VPN using the config file that I had just altered. Finally, the VPN connection has been initiated through the SSL/TLS tunnel. Figure 13 illustrates the overall configuration of the system.

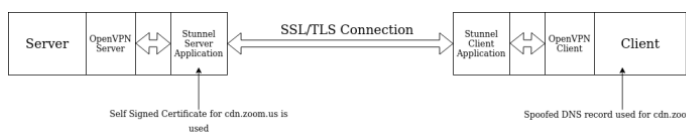

Figure 13. Stunnel and OpenVPN, Client Server Architecture Diagram

### III. RESULTS AND DISCUSSION

After completing all of the configurations mentioned in the previous sections, I tested the system by downloading a single file from a website, which is not included in the content-based package. If the configuration is working properly, the file should be downloaded with the content-based package data quota. The first approach I used was creating a fake web server without SSL. The server was working fine, and I could download files from the fake web server without any issues. But when checking the data usage, data had been deducted from the standard data quote instead of the content-based package data quote. It proves that the ISP filters network traffic not only by the hostname and the SNI, but also by the connection protocol. Standard HTTP connections are not allowed to use content-based packages.

The second approach was to configure SSL on the same web server. After configuring SSL with a self-signed certificate, I could download the same files I downloaded earlier. When checking the data usage, data had been deducted from the content-based package data quote. It confirms that the ISP filters network traffic by the hostname, the SNI, and the connection protocol. As the third approach, I installed and configured the Squid proxy server on the server, and then I connected the local computer to the proxy server through the Stunnel SSL/TSL tunnel. Then I could easily route network traffic of the local computer through the proxy server, and it used content-based package data quote. As the final approach, I configured the OpenVPN server on the remote server and connected the local computer to the VPN server through the Stunnel SSL/TLS connection. Then I had a fully configured VPN network available on the local computer, and I could route almost all the network traffic through the VPN. It consumes the content-based package data quote as expected.

### IV. CONCLUSION

Internet Service providers offer Content-based internet packages for a low cost. Most ISP's filter network traffic using the hostname, SNI, and connection protocol. When it comes to the hostname and SNI, Almost every ISP allows the root domain name and all the other subdomains for the particular service to use the content-based package data quote. For example, if we consider a content-based package that allows users to surf on Facebook, it will allow all the domains and subdomains such as,

- facebook.com
- www.facebook.com
- m.facebook.com
- mobile.facebook.com
- Abc.facebook.com
- fake-domain.facebook.com
- fake.domain.facebook.com

Almost all the ISPs only allow SSL Connections to consume Content-based internet packages data quotes. Therefore we can use any domain or subdomain related to the content-based package to trick the ISP. If we use an existing domain or subdomain, it will close the real connection to the particular domain or subdomain. Therefore we can use a domain or a subdomain that does not exist to use all the services without interruption. By using this way, users can easily access any resource available on the internet using their content-based internet package data quote. However, this method requires more expertise to configure all the required applications and servers. Using an HTTP Injector software will be easier than this method, but there are some disadvantages to using an HTTP Injector.

We need an SSL/TLS enabled SSH connection in order to use an HTTP injector. SSL/TLS enabled SSH account can be created for free using several websites available on the internet. But there are a lot of connection stability issues, speed issues, security issues, and privacy issues with these free SSH accounts. If we configure our own SSL/TLS enabled SSH account, it also requires expertise in the server configuration, but still, there might be privacy and security concerns with the HTTP injector application. Using the method we discussed in this paper, Users can use any resource available on the internet by using content-based package data quotes, assuring their privacy and security.

### REFERENCES

[1] Internet Society. 2022. An Overview of Internet Content Blocking | ISOC Internet Society. [online] Available at: <https://www.internetsociety.org/resources/doc/2017/internet-content-blocking/> [Accessed 16 May 2022].

[2] W. M. Shbair, T. Cholez, A. Goichot and I. Chrisment, "Efficiently bypassing SNI-based HTTPS filtering," 2015 IFIP/IEEE International Symposium on Integrated Network Management (IM), 2015, pp. 990-995, doi: 10.1109/INM.2015.7140423.

[3] stunnel.2022. stunnel. [online] Available at: <https://www.stunnel.org/docs.html> [Accessed 16 May 2022].

[4] Squid-cache.org. 2022. squid : Optimising Web Delivery. [online] Available at: <http://www.squid-cache.org/Intro/> [Accessed 16 May 2022].

[5] OpenVPN. 2022. Community Resources | OpenVPN. [online] Available at: <https://openvpn.net/community-resources/> [Accessed 16 May 2022].

[6] D. Storm, "What can your ISP really see and know about you?", Computerworld, 2022. [Online]. Available: https://www.computerworld.com/article/3043490/what-can-your-isp-really-see-and-know-about-you.html. [Accessed: 15- Oct- 2022].

[7] W. M. Shbair, T. Cholez, Jerome Francois, Isabelle Chrisment: "A Survey of HTTPS Traffic and Services Identification Approaches". [Online] Available: http://arxiv.org/abs/2008.08339'>arXiv:2008.08339.